\documentclass[conference]{IEEEtran}
 \pdfoutput=1
\usepackage{balance}
\usepackage{epsfig}
\usepackage{amsmath}
\usepackage{float}
\usepackage{amsfonts}
\usepackage{amssymb}
\usepackage{multicol}
\usepackage{subfig}
\usepackage{mathrsfs}
\usepackage{amsmath}
\usepackage{graphicx}
\usepackage{psfrag}
\usepackage{pdfsync}
\usepackage{booktabs}
\usepackage{color}
\usepackage{cite}
\usepackage{hyperref}
\usepackage{enumerate}
\usepackage{wrapfig}     

\usepackage{flushend}
\begin{document}
\title{Finite Length Analysis of LDPC Codes }

\author{\IEEEauthorblockN{Md. Noor-A-Rahim, Khoa D. Nguyen and Gottfried Lechner}
\IEEEauthorblockA{Institute for Telecommunications Research\\
University of South Australia\\
Adelaide, Australia\\
Email: noomy004@mymail.unisa.edu.au, \{khoa.nguyen, gottfried.lechner\}@unisa.edu.au}
}

\maketitle

\begin{abstract}
In this paper, we  study the performance of finite-length LDPC codes in the waterfall region. We propose an algorithm to predict the error performance of finite-length LDPC codes over various binary memoryless channels. Through numerical results, we find that our technique gives better performance prediction compared to existing techniques.
\end{abstract}

\section{Introduction}
Low-density parity-check (LDPC) codes are known to be the most popular coding technique due to their capacity achieving property. Moreover, LDPC codes offer low decoding complexity. Thus, analytical investigation of LDPC codes has a great importance. Asymptotic  behaviors of LDPC codes are well understood and the performance can be measured through the density evolution or EXIT chart analysis. In contrast of asymptotic analysis,  not much is known about the behavior of  finite-length LDPC codes.

Finite-length performance of LDPC codes is divided into two regions, namely error floor region and waterfall region. Error floor occurs due to small cycles generated in the code, where error performance of the code does not decrease as rapidly as it might be expected. On the other hand, in the waterfall region, error performance of the code drops off quickly as a function of channel parameter. Finite-length behavior of LDPC codes over binary erasure channel was first investigated in \cite{Stop_Set}, which describes both the waterfall and error floor regions. This approach was based on stopping set analysis. Although stopping set analysis gives an accurate estimation, it becomes impractical for large block length due to very high computational complexity. A similar approach with slightly lower complexity  was proposed in \cite{Left_Degree}. In \cite{Scaling}, scaling law method was introduced, which can predict the performance of finite-length LDPC codes almost accurately. The prediction accuracy of scaling law method depends on how accurately one can find  the scaling parameters and decoding threshold. Although scaling law method provides a low complexity analysis, findings of scaling parameters are not easy task for all ensembles and decoding algorithms. More recently, another approach  was proposed in \cite{Raman}, which provides finite-length performance estimation in the waterfall region. This approach  models the channel parameters as random variables and approximates their probability distribution functions.  Considering the decoding failure as an event when the realized channel quality is worse than the decoding threshold, block error probability was obtained. Using the extrinsic information transfer (EXIT) chart analysis and the obtained block error probability, this method calculates the bit error probability.  This threshold method offers an efficient and low complexity estimation in the waterfall region than the above  mentioned techniques. However, it gives a poor prediction for small block lengths.

In this paper, we develop a new method to investigate the finite-length behavior of LDPC codes. Utilizing the density evolution technique \cite{DE}, our approach can predict the bit error probability of any given ensemble. This approach also holds for different memoryless channels and different decoding algorithms. Through comparison with  simulation results, we show that our approach gives a close approximation for LDPC bit error performance. Although our proposed method exhibits higher computational complexity, it provides more accurate estimations than that obtained from the method proposed in  \cite{Raman}.

The rest of the paper is organized as follows. In Section II, we present some background on LDPC codes and memoryless channel models. The proposed approach to estimate the bit error probability is introduced in Section III.  In Section IV, we present the numerical results. Finally, we conclude our paper in Section V.

\section{Background}

\subsection{LDPC Codes}
LDPC codes were first introduced by Robert Gallager  in his PhD dissertation \cite{GallagerLDPC}. In general, an LDPC code is represented by a bipartite graph consisting of $N$ variable nodes and $M$ check nodes. Each variable node represents a coded bit and each check node represents a check equation indicating that the connecting variable nodes sum to zero. An LDPC code is called regular $(d_v,d_c)$, if every variable node has exactly $d_v$ edges and every check node is connected via exactly $d_c$ edges, where $d_v$ and $d_c$ are known as the degree of the variable and check nodes, respectively. On the other hand, irregular LDPC codes are those that have non-uniform row degrees and non-uniform column degrees. In general, the degree distributions are characterized by the following functions \cite{SPM685}:
\begin{itemize}
 \item $\mathit{v}(x) = \sum\limits_{i=2}^{d_{v_{\text{max}}}} \mathit{v}_i x^{i} $,  where $\mathit{v}_i$ is the fraction of variable nodes of degree $i$.
  \item $\mathit{h}(x) = \sum\limits_{i=2}^{d_{c_{\text{max}}}} \mathit{h}_i x^{i} $,  where $\mathit{h}_i$ is the fraction of check nodes of degree $i$.
  \item $\lambda(x) = \sum\limits_{i=2}^{d_{v_{\text{max}}}} \lambda_i x^{i-1} $, where $\lambda_i$ is the fraction of edges that are connected to degree $i$ variable nodes.
  \item $\rho(x) = \sum\limits_{i=2}^{d_{c_{\text{max}}}} \rho_i x^{i-1} $, where $\rho_i$ is the fraction of edges that are connected to degree $i$ check nodes,
 \end{itemize}
where $d_{v_{\text{max}}}$ and $d_{c_{\text{max}}}$ are the maximum number of degree of variable nodes and check nodes, respectively. For the above mentioned degree distributions, the code rate becomes \cite{SPM685},
\begin{align}
   R = 1-\frac{\sum\limits_{i}\mathit{v}_ii}{\sum\limits_{i}\mathit{h}_ii} = 1-\frac{\sum\limits_{j}\rho_j/j}{\sum\limits_{j}\lambda_j/j}\nonumber
\end{align}
\subsection{Channel Models}
In this paper, we consider  memoryless noisy channels. In general, a memoryless channel can be defined by input-output transition probabilities $p(y|x)$ for $x \in \mathcal{X}$ and $y \in \mathcal{Y}$, where $\mathcal{X}$ and $\mathcal{Y}$ are the input and output alphabets, correspondingly. We estimate the finite-length performance of LDPC codes over the following channels:

\subsubsection{Binary Erasure Channel (BEC)}
A binary erasure channel (BEC) with erasure probability $\epsilon$ implies $\mathcal{X}=\{0,1\}$ and $\mathcal{Y}=\{0,1,?\}$,  while $p(y = ?|x =0) = p(y = ?|x =1) = \epsilon$ and  $p(y = 0|x =0) = p(y = 1|x =1) = 1- \epsilon$, where ? indicates an erasure.

\subsubsection{Binary Symmetric Channel (BSC)}
In binary symmetric channel (BSC), we define channel input and channel output alphabets as $\mathcal{X}=\{0,1\}$ and $\mathcal{Y}=\{0,1\}$, respectively, where the channel flips the transmitted bits with a probability $\epsilon$. Thus, a received bit either in error with probability $\epsilon$ or received correctly with probability $1-\epsilon$. Mathematically, $p(y = 1|x =0) = p(y = 0|x =1) = \epsilon$ and  $p(y = 0|x =0) = p(y = 1|x =1) = 1- \epsilon$. For BSC, $\epsilon$ is called crossover probability.

\subsubsection{Additive White Gaussian Noise (AWGN) Channel}
The channel  output of binary input additive white Gaussian noise (BIAWGN) channel can be described by $y = x+z$, where $x\in \{-1,+1\}$ and $z$ is the Gaussian random noise with zero mean and variance $(\sigma_n^2)$. Thus, the transition probability of AWGN channel becomes:
\begin{align}
   p(y|x) = \frac{1}{\sqrt{2\pi\sigma_n^2}}e^{-\frac{(y-x)^2}{2\sigma_n^2}}\nonumber
\end{align}
The channel parameter of AWGN channel is expressed as the ratio of the energy per bit $E_b$ and the noise power spectral density $N_0$, which is also known as signal-to-noise ratio (SNR) \cite{SPM685}. We refer the channel log-likelihood ratio (LLR) as $L$, measured by $L_i = \log\frac{p(x_i=+1|y_i)}{p(x_i=-1|y_i)}$. Further analysis shows that the received LLR over BIAWGN channel can be obtained from $L_i = \frac{2}{\sigma_n^2} y_i$. We denote $(p_c)$ as the probability of receiving a bit in error due to the  AWGN channel. With a symmetric Gaussian distribution, the error probability $(p_c)$ can be calculated by $p_c = Q\big(\frac{1}{\sigma_n}\big)$ \cite{Raman}.

\subsection{Decoding Techniques}
The decoding algorithms used to decode the LDPC codes are  called message passing algorithms. As the name suggests, these algorithms are associated with passing messages from variable nodes to check nodes and vice-versa. Depending on the algorithms, the type of messages that passed between the nodes are different. In this paper, we consider erasure decoding algorithm for BEC \cite{SPM685}, Gallager A decoding algorithm for BSC  \cite{GallagerLDPC}  and  belief propagation decoding algorithm for AWGN channel \cite{SPM685}.

\subsection{Density Evolution}
Density evolution (DE) is an analytical tool to analyze the performance of iterative decoding of a particular code ensemble. This technique tracks the evolution of probability density functions of the messages through the decoding process, which provides the convergence calculation for a given code ensemble \cite{SPM685}. The code converges when the probability density function (pdf) of the message converge to some desired distribution and the convergence/decoding threshold is the lowest/worst channel parameter such that the decoder converges.  Density Evolution on the BEC is straight forward. For a $(d_v,d_c)$ regular ensemble the updates of variable node decoder and check node decoder are obtained by the following equations \cite{SPM685}:
\begin{description}
 \item[Variable node update:]
  \begin{align}
   m_{v}^{(l)}= \epsilon(m_{c}^{(l-1)})^{d_v-1}\nonumber
   \end{align}
 \item[Check node update:]
\begin{align}
  m_{c}^{(l)}= 1-(1-m_{v}^{(l)})^{d_c-1},\nonumber
\end{align}
\end{description}

where $m_v^{(l)}$ and $m_c^{(l)}$  are the erasure probabilities of outgoing messages from variable node and check node, respectively at iteration $l$. The above recursions are termed as density evolution equations, since they describe how the erasure probability of the iterative decoder evolves as a function of the iteration number $l$. Although for BEC, density evolution is equivalent to tracking the erasure probability, in general tracking the evolution of probability density functions is required.

\section{Proposed Algorithm}
In the asymptotic density evolution, the realized channel-parameter observed by each codeword is concentrated and fixed. However, for finite-length codes, the channel observed by a codeword varies. Similar to \cite{Raman}, our approximation technique takes into account this randomness of the channel to evaluate the performance of finite-length codes, while assuming that the codes are cycle-free. For example, in erasure channel, we  define a random variable $E$, the obtained  bit erasure probability, which has a binomial distribution with success probability $\epsilon$. For large $N$,  this distribution can be approximated by a Gaussian distribution $\mathcal{N}(\mu,\sigma^2)$,  where $\mu = \epsilon$ and  $\sigma^2= \frac{\epsilon(1-\epsilon)}{N}$. Conditioning on a channel realization (erasure/error probability), we first find the conditional decoded erasure/error probability from density evolution. Then by marginalizing over the channel realizations, we find the erasure/error probability.
Let $k$ denotes the  realized number of erasures/crossings or the realized empirical average LLR. Conditioned on $k$, we consider $m_{\text{ch}|k}$, $m_{v|k}^{(l)}$ and $m_{c|k}^{(l)}$ as the mean of messages obtained from the channel, variable node decoder and check node decoder, respectively at iteration $l$.  We denote $e_k$  as the realized channel erasure/error probability and $P_{b|k}$ as the erasure/error probability returned by the decoder given $k$ erasures/crossovers/LLR realized at the channel.   Using the following steps we can estimate the  performance of finite-length LDPC codes.

\begin{description}
\item[\textbf{Quantization:}]\hskip 8ex Quantize the realized channel erasures/crossovers (BEC/BSC) or realized  channel LLR (AWGN channel).
\vskip 1.5ex
  \item[\textbf{$P_{b|k}$ calculation:}]\hskip 12ex For each of the quantized value $k$, find the corresponding erasure/error probability $P_{b|k}$ returned by the decoder using following steps:
\vskip 1ex
  \begin{enumerate}
    \item At the variable node decoder, compute $m_{v|k}^{(l)}$  for given $m_{\text{ch}|k}$ and $m_{c|k}^{(l-1)}$.
\vskip .8ex
  \item At the check node decoder, compute $m_{c|k}^{(l)}$ for given $m_{v|k}^{(l)}$.
\vskip .8ex
  \item Repeat step 1 and 2 for a maximum number of iterations $l_{\text{max}}$.
\vskip .8ex
  \item Find the conditional erasure/error probability  $P_{b|k}$ using $m_{\text{ch}|k}$ and $m_{c|k}^{(l_{\text{max}})}$ [\autoref{bec_c}, \autoref{bsc_c} and \autoref{awgn_c}].
 \end{enumerate}
\vskip 1.5ex
  \item[ \textbf{Marginalization:}]\hskip 11ex Marginalize $P_{b|k}$ over $k$ to get the erasure/error probability $(P_b)$.
  \begin{align}\label{eq_5}
 P_b= \sum\limits_k P_r(e_k) P_{b|k},
   \end{align}
\end{description}
where $P_r(e_k)$ denotes the probability of having erasure/error probability $(e_k)$, which can be obtained from the distribution of the realized channel erasure/error probability as mentioned earlier. The rules of finding  $m_{v|k}^{(l)}$,  $m_{c|k}^{(l)}$ and $P_{b|k}$  depend on the variation of channel type and decoding algorithm.

\section{Numerical Results}
In this section, we present the estimation results for  LDPC codes according to the algorithm described in Section III. We compare our estimation results with the simulation results and threshold method \cite{Raman} results for different memoryless channels.

\subsection{On the Binary Erasure Channel}
For BEC, the findings of variable node update, check node update and conditional erasure probabilities are as follows:
\begin{description}
  \item[Variable node update:]
  \begin{align}
   m_{v|k}^{(l)}= m_{{\text{ch}}|k}\lambda(m_{c|k}^{(l-1)}).\nonumber
   \end{align}
 \item[Check node update:]
\begin{align}
  m_{c|k}^{(l)}= 1-\rho(1-m_{v|k}^{(l)}).\nonumber
\end{align}
\item[Conditional erasure probability:]
\begin{align} \label{bec_c}
  P_{b|k}= m_{{\text{ch}}|k}\mathit{v}(m_{c|k}^{(l_{\text{max}})})
\end{align}
\end{description}

\begin{figure}[htbp]
   \centering
  \includegraphics{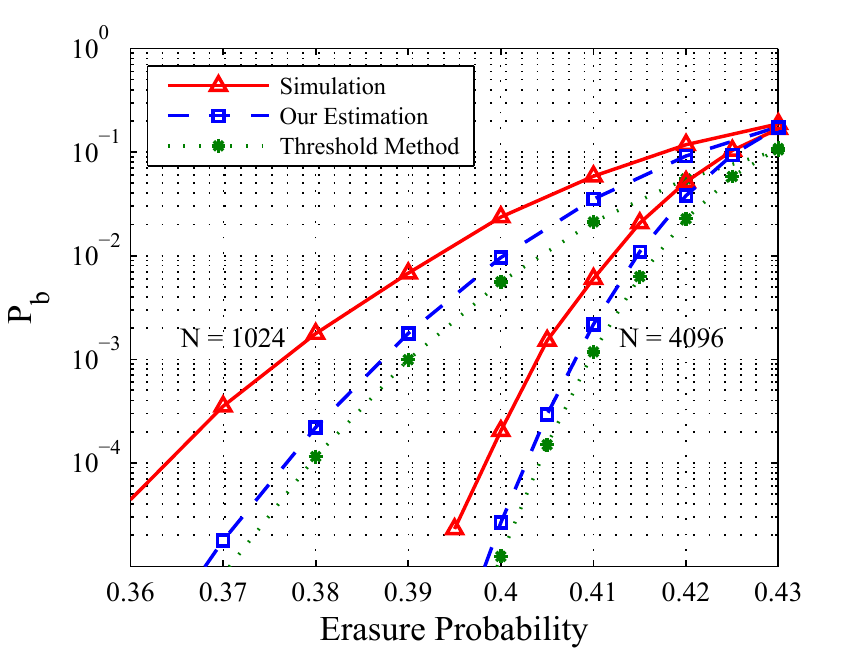}\\
  \caption{Erasure Probability  comparison for $(3,6)$-regular LDPC Code between the simulation results, our estimations and threshold method results proposed in \cite{Raman}.}\label{becR}
\end{figure}

For $k$ number of erasures from channel, we calculate the observed erasure probability by $e_k = \frac{k}{N}$. We find $P_r(e_k)$ from the pdf of realized channel erasure probability and marginalize $P_{b|k}$ over  $k$ using \autoref{eq_5} to calculate the erasure probability $(P_b)$ of a given ensemble. \autoref{becR} shows comparison between the simulation result and our prediction  method for $(3,6)$-regular LDPC codes.

\subsection{Over Binary Symmetric Channel}
\begin{figure}[htbp]
   \centering
  \includegraphics{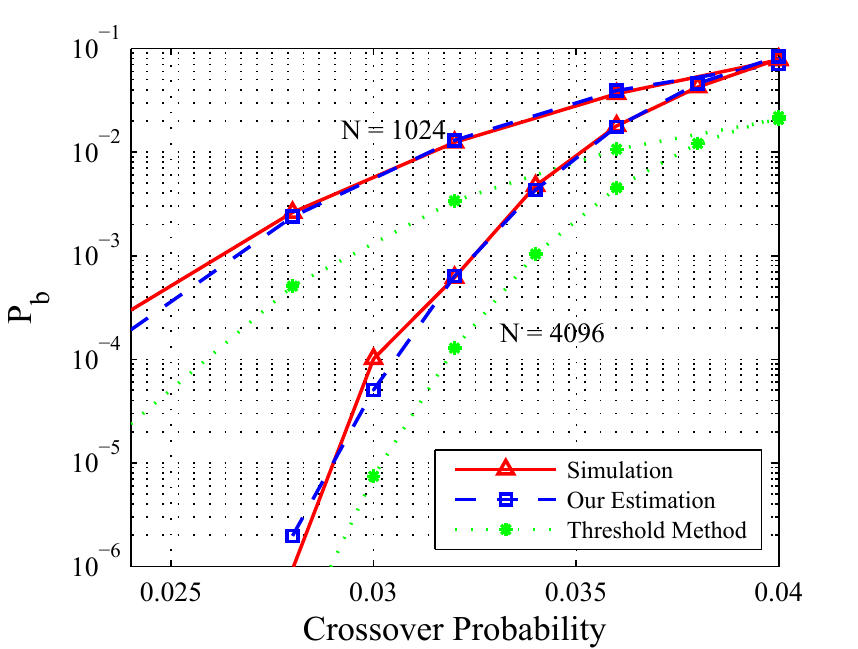}\\
  \caption{Bit error rate comparison between the simulation results and analytical results for (3,6)-regular LDPC Codes over BSC under Gallager A algorithm.}\label{bscR}
\end{figure}

On the binary symmetric channel (BSC), we consider Gallager A decoding  algorithm \cite{GallagerLDPC}, which provides an easy analysis similar to the BEC. In this algorithm, the message passing between variable node and check node occurs in the following manner \cite{Amin}. At each iteration, a variable node $v$ sends a value $b \in \{0, 1\}$ to a check node $c$, when all the incoming messages from neighboring check nodes other than check node $c$ are $b$; otherwise $v$ sends its received value from channel to $c$. On the other hand, a check node $c$ sends to variable node $v$ the sum (mod 2) of all incoming messages from neighboring variable nodes other than $v$. Thus, we calculate $m_{v|k}^{(l)}$ and $m_{c|k}^{(l)}$ from the following equations \cite{GallagerLDPC},
\begin{description}
  \item[Variable node update:]
  \begin{align}
   m_{v|k}^{(l)}= &(1-m_{{\text{ch}}|k})\lambda(m_{c|k}^{(l-1)}) +  \nonumber\\
   & m_{{\text{ch}}|k}  \big(1-\lambda(1-m_{c|k}^{(l-1)})\big). \nonumber
   \end{align}

 \item[Check node update:]
  \begin{align}
   m_{c|k}^{(l)}= \frac{1-\rho(1-2m_{v|k}^{(l)})}{2}.\nonumber
  \end{align}
\end{description}
After a maximum number of iteration, the conditional error probability becomes,
\begin{align}\label{bsc_c}
 P_{b|k} = &(1-m_{{\text{ch}}|k})\mathit{v}(m_{c|k}^{(l_{\text{max}})}) +  \nonumber\\
   & m_{{\text{ch}}|k} \big(1-\mathit{v}(1-m_{c|k}^{(l_{\text{max}})})\big)
\end{align}
For $k$ number of crossover from channel, we calculate the observed crossover probability by $e_k = \frac{k}{N}$. We find $P_r(e_k)$ from the pdf of realized channel crossover probability and marginalize $P_{b|k}$ over  $k$ using \autoref{eq_5} to calculate the crossover probability $(P_b)$ of a given ensemble. For different block lengths $N$, \autoref{bscR} shows the comparison between our predictions and the simulation results for $(3,6)$-regular LDPC code under Gallager A decoding algorithm.

\subsection{Over AWGN Channel}
\begin{figure}[htbp]
   \centering
  \includegraphics{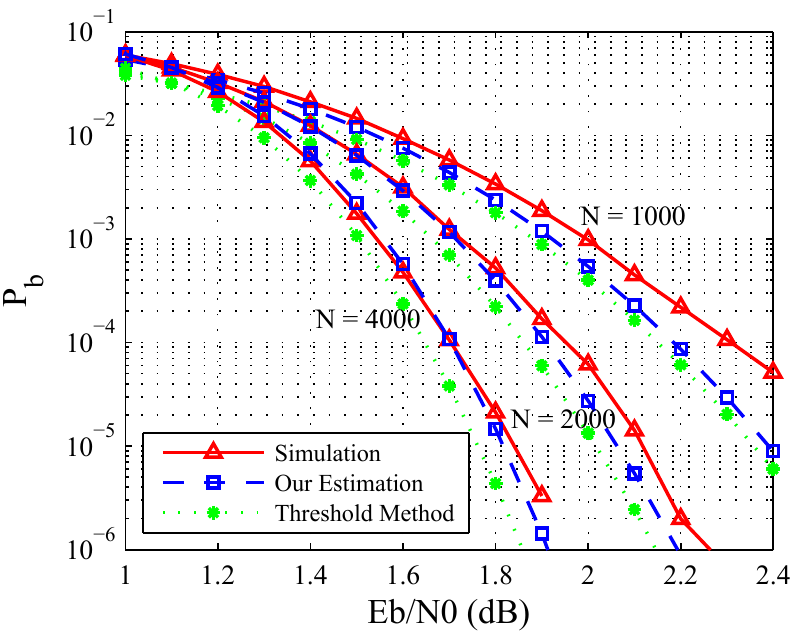}\\
  \caption{Bit error rate comparison between the simulation results, our estimation and threshold method \cite{Raman} for $(3,6)$-regular LDPC Code over AWGN channel.}\label{awgnR}
\end{figure}

\begin{figure}[htbp]
   \centering
  \includegraphics{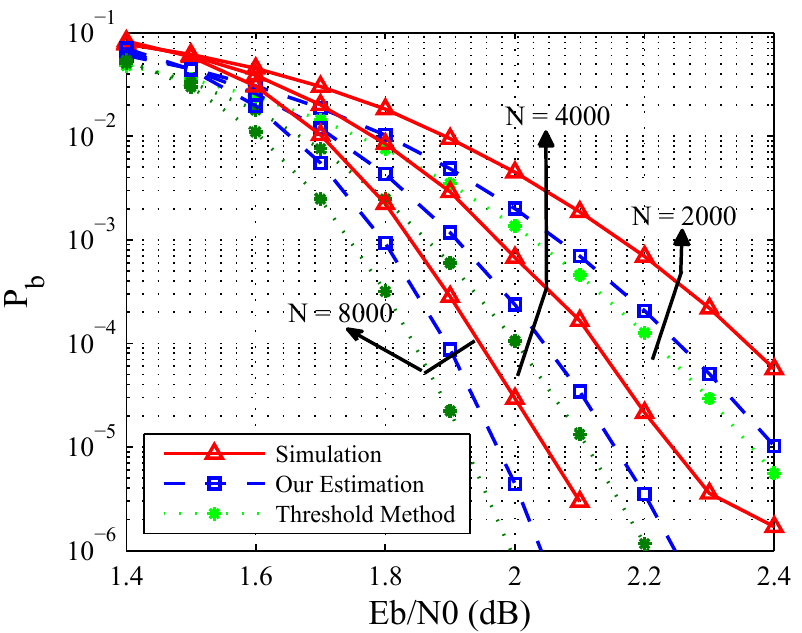}\\
  \caption{ Bit error rate comparison between the simulation, our estimation and threshold method \cite{Raman} for irregular LDPC codes. We specified the degree distribution pair of the code as $\lambda(x) = 0.4x^2 + 0.4 x^5 + 0.2x^8$ and $\rho(x) = x^8$}\label{awgnIR}
\end{figure}

For AWGN channel, we consider LDPC codes under belief propagation decoding algorithm. To implement the density evolution based on belief propagation algorithm, we have utilized Gaussian approximation technique showed in \cite{GA}, which provides a close result compare to the actual implementation. Similar to \cite{GA}, for a Gaussian LLR message with $\mathcal{N}(\mu,\sigma^2)$, we consider symmetric condition $\sigma^2 = 2\mu$. We also define
\begin{align}
\nonumber
\end{align}
 \begin{align}
    \phi(x)=
  \begin{cases}
   1 - \frac{1}{\sqrt{4\pi x}}\int_{-\infty}^{\infty} \big(\tanh \frac{u}{2}\big) e^{-\frac{(u-x)^2}{4x}}du, &  \text{if} \hspace{0.2 cm} x>0 \\\nonumber
   1, &  \text{if} \hspace{0.2 cm} x=0 \nonumber
  \end{cases}
 \end{align}
Using the above approximation for AWGN channel, we track the mean of LLRs that passing between variable node and check node decoder.
\begin{description}
\vskip 1ex
  \item[Variable node update:]
  \begin{align}
   m_{v|k}^{(l)}=\sum\limits_{i=2}^{d_{v_{\text{max}}}} \lambda_i\big( m_{{\text{ch}}|k} + (i-1)m_{c|k}^{(l-1)}\big).\nonumber
   \end{align}
 \item[Check node update:]
  \begin{align}
  m_{c|k}^{(l)}=&\sum\limits_{j=2}^{d_{c_{\text{max}}}}\rho_j \phi^{-1}\Big(1-\big[1-\sum\limits_{i=2}^{d_{v_{\text{max}}}}  \lambda_i\phi( m_{{\text{ch}}|k} + \nonumber\\
  &(i-1)m_{c|k}^{(l-1)})\big]^{j-1}\Big). \nonumber
  \end{align}
\item[Conditional error probability:]
\begin{align}\label{awgn_c}
   P_{b|k}= \sum\limits_{i=2}^{d_{v_{\text{max}}}} \mathit{v}_iQ\left(\sqrt{\frac{ m_{{\text{ch}}|k} + im_{c|k}^{(l_{\text{max}})}}{2}}\right)
\end{align}
\end{description}
Each quantized LLR value $k$ corresponds to the error probability given by $e_k = Q(\sqrt{\frac{k}{2}})$. Recall that, with noise variance $\sigma_n^2$ obtained from signal to noise ratio, the average error probability from channel $p_c$ is given by $p_c = Q(\frac{1}{\sigma_n})$. For finite-length $N$, the distribution of $e_k$ can be approximated by Gaussian distribution with $\mathcal{N}(p_c,\frac{p_c(1-p_c)}{N})$ and we calculate $P_r(e_k)$ from this distribution. Then by marginalizing $P_{b|k}$ over  $k$ using \autoref{eq_5}, we can obtain the error probability $(P_b)$ of a given ensemble. In \autoref{awgnR}, our estimations for $(3,6)$-regular LDPC codes of different block lengths are compared to the simulation and reference \cite{Raman} results. Then we compare our estimation with the simulation result for irregular case. For a rate $\frac{1}{2}$ irregular code,  \autoref{awgnIR} shows the comparison between our estimation, simulation and reference \cite{Raman} results.

\subsection{Discussions}
From the above comparisons, we find our estimation technique better than the method in \cite{Raman}.  Similar to \cite{Raman}, we observe the reduction in gap between estimation and simulation results for large block length. We also observe that, for channel parameters close to the decoding threshold of the code, our proposed algorithm provides very good estimations. Moreover, our proposed method is useful in the case, where it is difficult to find the decoding threshold accurately. A suitable example of such case can be the anytime spatially coupled code proposed in \cite{our_anytime}, where it is hard to find the decoding threshold at a certain delay. Thus, we can use our proposed algorithm to estimate  the finite-length performance of such codes.   It is worth to mention that our proposed method can also be applicable to estimate the finite-length performance of repeat accumulate (RA) codes in the waterfall region.

\section{Conclusion}
In this paper, we present an algorithm to estimate the finite-length behavior of LDPC codes in the waterfall region.  From the comparison with the simulation results, we find our algorithm as a good prediction method over different channels without knowing the threshold or any other  parameters. It will be interesting to see how this method can be used to optimize LDPC codes.


\bibliography{WCNC14}
\bibliographystyle{IEEEtran}

\end{document}